\documentclass[twocolumn,showpacs,preprintnumbers,amsmath,amssymb,prl]{revtex4}

\usepackage{graphicx}            
\usepackage{dcolumn}             
\usepackage{bm}                  

\newcommand{ \lczmo } {La$_{2}$Cu$_{\rm 1-p}$(Zn,Mg)$_{\rm p}$O$_{4}$}

\begin{document}

\topmargin = -1cm

\preprint{APS/123-QED}

\title{Quantum vs. Geometric Disorder in a
Two-Dimensional Heisenberg Antiferromagnet}

\author{O.P. Vajk}
\affiliation{
Department of Physics, Stanford University, Stanford, California 94305
}
\author{M. Greven}
\affiliation{Department of Applied Physics and Stanford
Synchrotron Radiation Laboratory, Stanford, California 94305 }

\date{\today}       

\begin{abstract}
We present a numerical study of the spin-1/2 bilayer Heisenberg
antiferromagnet with random inter-layer dimer dilution.
From the temperature dependence of the uniform susceptibility
and a scaling analysis of the spin correlation length we deduce
the ground state phase diagram as a function of nonmagnetic
impurity concentration $p$ and bilayer coupling $g$. At the
site percolation threshold, there exists a multicritical point
at small but nonzero bilayer coupling $g_m = 0.15(3)$.
The magnetic properties of the single-layer material
\lczmo~near the percolation threshold appear to be controlled
by the proximity to this new quantum critical point.
\end{abstract}


\maketitle

Quantum phase transitions (QPTs) in the presence of disorder are
the subject of considerable current interest as they exhibit rich
new physics, but are rather poorly understood.
Experimental examples include
the cuprate superconductors \cite{attfield98},
heavy fermion compounds \cite{HFC},
metal-insulator \cite{MIT} and
superconductor-insulator transitions \cite{SIT}, quantum Hall
effect \cite{QHE},
and quantum magnets
\cite{Wu}. A significant amount of
theoretical and numerical work has been devoted to the study of
the random Ising chain in a transverse field \cite{1DIsing},
the simplest quantum many-body system with quenched disorder,
and to its
analog in higher dimensions
\cite{2DIsing,senthil96}. Partly because of its
relevance to cuprate superconductivity, the spin-1/2 square-lattice
Heisenberg antiferromagnet (SLHAF) has attracted enormous
interest \cite{greven95}. While the effects of a single nonmagnetic impurity are
well understood \cite{dilute}, there exist few theoretical results
for finite impurity concentrations \cite{chernyshev01}. Given the
considerable challenges, theory for quantum systems with disorder
is often guided by insight provided by numerical work.

In the absence of disorder, the nearest-neighbor (NN)
SLHAF can be driven through a QPT
by introducing a parameter analogous to the
transverse field for the Ising model.
For example, this is achieved by introducing frustrating next-NN couplings
\cite{chakravarty89} or
by coupling two square lattices to form an antiferromagnetic
bilayer \cite{shevchenko00}.
It had been argued that random site or bond dilution of the SLHAF in the extreme
quantum limit of spin-1/2 may lead to a non-trivial QPT
\cite{senthil96,site+bond,kato00}.
However, recent experimental \cite{vajk02} and numerical
\cite{vajk02,sandvik02a}
work suggests that the ground state remains ordered for non-magnetic
impurity concentrations $p$ up to the site percolation threshold
$p_*$, and that the critical cluster at $p = p_*$ appears to
have a nonzero staggered moment
\cite{sandvik02a}, which
would imply that the percolation transition is classical.

In this Letter, we present numerical results for the site-diluted
spin-1/2 NN bilayer antiferromagnet, with disorder
that is fully correlated between the layers (``dimer" dilution).
At zero bilayer coupling, this problem reduces to the previously
studied diluted
spin-1/2 SLHAF. By increasing the strength of the bilayer coupling
from zero, we
are able to increase quantum fluctuations beyond those for the
spin-1/2 SLHAF. Our finite-temperature results allow us to extract the
ground state phase diagram as a function of the strength of
quantum fluctuations and the degree of disorder,
and they reveal a new multicritical point
at $p = p_*$ for small, but nonzero bilayer coupling.
This phase diagram resembles that of the
diluted two-dimensional Ising ferromagnet in a transverse
field \cite{senthil96}. The present results are relevant to recent
experimental findings for the model spin-1/2 SLHAF \lczmo~\cite{vajk02}.

We study the Hamiltonian
\begin{eqnarray}
H = J \sum_{\langle i,j \rangle, n = 1,2 } \ \epsilon_i\epsilon_j
{\bf S}_{i,n} \cdot{\bf S}_{j,n} + J_{\perp} \sum_{i}
\epsilon_i{\bf S}_{i,1} \cdot{\bf S}_{i,2} \label{hamiltonian}
\end{eqnarray}
where $J$ ($J_{\perp}$) is the antiferromagnetic planar (bilayer)
coupling, the first sum is over all planar NN pairs,
and $\epsilon_i = 0$ ($\epsilon_i = 1$) with probability $p$
($1-p$). We define the
reduced bilayer coupling $g \equiv J_{\perp}/J$, and work with
units in which $J=a=k_B = g\mu_B = \hbar = 1$ ($a$ is the
lattice constant). If a site is removed on one layer, the
corresponding site in the adjacent layer is also removed. This
constraint preserves the percolation properties of the square
lattice. It also ensures that
there are no unpaired ``dangling" spins. Unlike
for the diluted SLHAF ($g=0$), the uniform susceptibility
$\chi_u$ for the bilayer then contains no Curie-like term.
The temperature dependence of
$\chi_u$ can then be used to determine the critical bilayer
coupling $g_c (p)$
for the diluted system. We use the loop-cluster
Monte Carlo method \cite{wiese94}, which has provided good
results for the diluted spin-1/2 ladder \cite{greven98}
and SLHAF \cite{vajk02}.
Simulations are performed
on large lattices of up to $N = 512 \times 512 \times 2$ sites,
to temperatures as low as $T = 1/300$, with Trotter numbers of $20/T$,
and by averaging 10 to 100 random configurations.  By
keeping the planar lattice size significantly larger than the
correlation length $\xi (g,p,T)$, we are
able to avoid finite-size effects.
Several concentrations are chosen, spanning the range
from the pure system up to $p=0.5$, beyond the site
percolation threshold $p_*=0.40725379(13)$ \cite{newman00}. In
addition to $\chi_u (g,p,T)$ and $\xi (g,p,T)$, we also compute
the staggered susceptibility $\chi_{st} (g,p,T)$.  The
susceptibilities are defined as
$N T \chi_u =  \left[ \left< (\sum_i S_i^z )^2 \right> \ \right]$ {\rm and}
\ $
N T \chi_{st}= \left[ \left< (\sum_i (-1)^i \ S_i^z )^2 \right> \right]$,
where the sum is over all sites $N$ and $\left[ ... \right]$
indicates the disorder average. The instantaneous staggered
correlation function is given by
$NC({\bf r}) = sign({\bf r}) \left[ \left< \sum_i S_i^z S_{i+{\bf r}}^z
\right> \right]$,
where $sign({\bf r}) =$ 1 (-1) for ${\bf r}$ separating sites on
the same (different) sublattice. The correlation length is
obtained from the behavior $C({\bf r}) \sim e^{-|{\bf r}|/\xi}$ at large
distances.

\begin{figure}
\includegraphics[width=8.5cm]{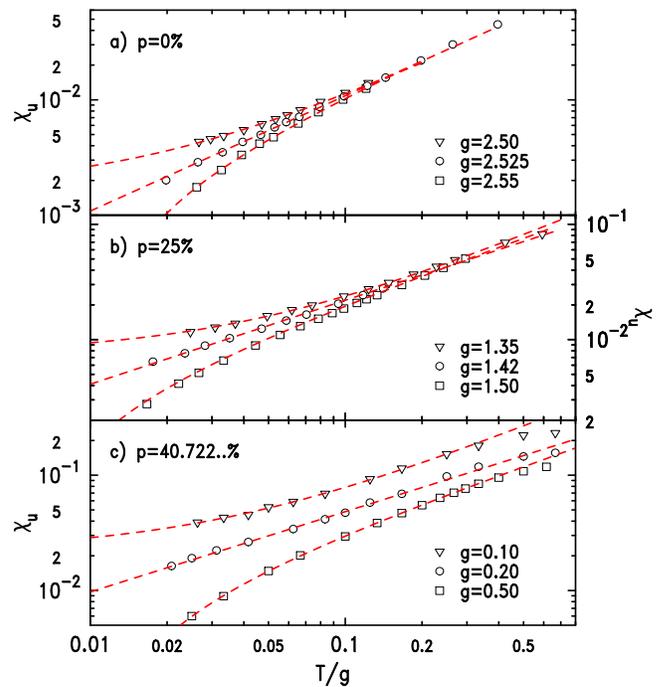}
\caption{\label{chi_u}
    Temperature dependence of the uniform susceptibility.
        (a) Pure bilayer: $g_c \approx 2.525$ and
        $\chi_u (g_c,0,T)  \sim T$,
        as found previously \cite{shevchenko00}.
        (b) $p=0.25$: $g_c \approx 1.42$ and $\chi_u \sim T^{0.73(5)}$.
        (c) Just below the percolation threshold: $g_c$ is significantly
        reduced, and $\chi_u \sim T^{0.70(7)}$.
        }
\end{figure}

At the critical coupling of the pure bilayer, $\chi_u
(g_c,0,T) \sim T$ \cite{shevchenko00}. In the quantum disordered
phase ($g > g_c$), the strong bilayer coupling leads to a nonzero
singlet-triplet gap, which prevents long-range order in the ground
state, and $\chi_u$ is exponentially small at low temperature. In
the ordered phase ($g<g_c$), $\chi_u$ approaches a nonzero value
proportional to the spin stiffness \cite{chubukov94}. These
different characteristic behaviors have allowed the determination
of the critical bilayer coupling for the pure system from
finite-$T$ numerics \cite{shevchenko00}. This should also be possible
at nonzero $p$, as long as there
exists an ordered phase.
Figure \ref{chi_u} shows some of our results for $\chi_u$ for the
pure system, at the intermediate concentration of $p=25\%$, and at
a concentration just below the percolation threshold. Note that
$T$ is scaled by $g$, so that results for different concentrations
fall into approximately the same horizontal range.
Fitting the low-$T$ data to $\chi_u = c_1+ c_2 T^{\lambda}$
gives excellent results, as shown
by the dashed lines in Fig. \ref{chi_u}. We interpret positive
(negative) values of $c_1$ as indicative of an ordered
(quantum disordered) ground state.
For $p =0$, we
find $g_c \approx 2.525$, in very good agreement with
$g_c=2.525(2)$ obtained previously \cite{shevchenko00}.
The critical coupling decreases with increasing $p$, and
$\lambda$ decreases from its value of 1 for the pure system to
around 0.7 near the percolation threshold.
We conclude that order persists at $g=0.10$
even at $p=417/1024$ (40.722...\%),
very close to $p=p_*$ (40.725...\%).

In the ordered phase, the low-$T$ correlation length
is expected to be exponential in inverse temperature
as for the pure SLHAF \cite{chernyshev01,vajk02}, while
$\xi \sim T^{-1/z}$ at the quantum critical point \cite{sachdev92},
where $z$ is the dynamic critical exponent.
The crossover temperature $T_{dev}$ at which $\xi$ deviates from
a power law scales with the deviation from the critical
point as $T_{dev} \sim |g-g_c|^{\phi}$ ($p$ can be substituted for
$g$ for cuts at fixed $g$).  Similarly, in the
disordered phase, the low-$T$ correlation length
approaches a constant value, but follows power-law behavior at higher
temperatures.  Figure 2 shows scaling plots for
four cuts across the phase boundary.
For the pure bilayer, Fig. 2(a), for which the low-$T$ physics can be
mapped to the quantum nonlinear $\sigma$ model \cite{chakravarty89},
the Euclidean
time direction is equivalent to the spatial dimensions (Lorentz
invariance) and $z = 1$. At zero temperature, Euclidean time
extends to infinity, so that the critical exponents of the pure
system are those of the {\it three-dimensional
classical} Heisenberg model, for which
$\nu \approx 0.705$ and $\gamma \approx 1.39$ \cite{guida98}.
Since $\nu = \phi/z$, we therefore have fixed
$\phi=0.705$ in Fig. 2(a). We obtain excellent scaling
with $g_c = 2.5215(10)$, slightly lower than the value
obtained from Fig. 1(a) and in previous work
\cite{shevchenko00}. A similar result is obtained for
$\chi_{st}$ (not shown) using $\gamma = 1.39$.

At $p=25\%$ [Fig. 2(b)], we find $z = 1.07(2)$ and $\phi = 0.95(5)$, as
well as $g_c (p=0.25)=1.412(6)$, consistent with Fig. 1(b).
Scaling as a function of $p$ at fixed $g=1.42$ (not shown) gives
good results using the same exponents and
$p_c(g=1.42) = 24.85(15)\%$.
Near $p=p_*$, scaling as a function of $p$ is easier to obtain,
and our results for $g=0.20$ and $g=0$
(single layer) are shown in Figs. 2 (c) and (d), respectively. For
$g=0.20$, the exponents are $z = 1.3(1)$, $\phi = 1.45(10)$, and
$p_c = 39.9(3)\%$. Recent neutron scattering results for
\lczmo~\cite{vajk02} as well as numerical work
\cite{kato00,vajk02,sandvik02a} indicate that $p_c=p_*$ for the
spin-1/2 SLHAF. Therefore, at $g=0$, we fixed $p_c = p_*$. As is evident
from Fig. 2(d), this gives excellent scaling, with $z =
1.65(5)$ and $\phi = 1.8(1)$.

\begin{figure}
\includegraphics[width=8.5cm]{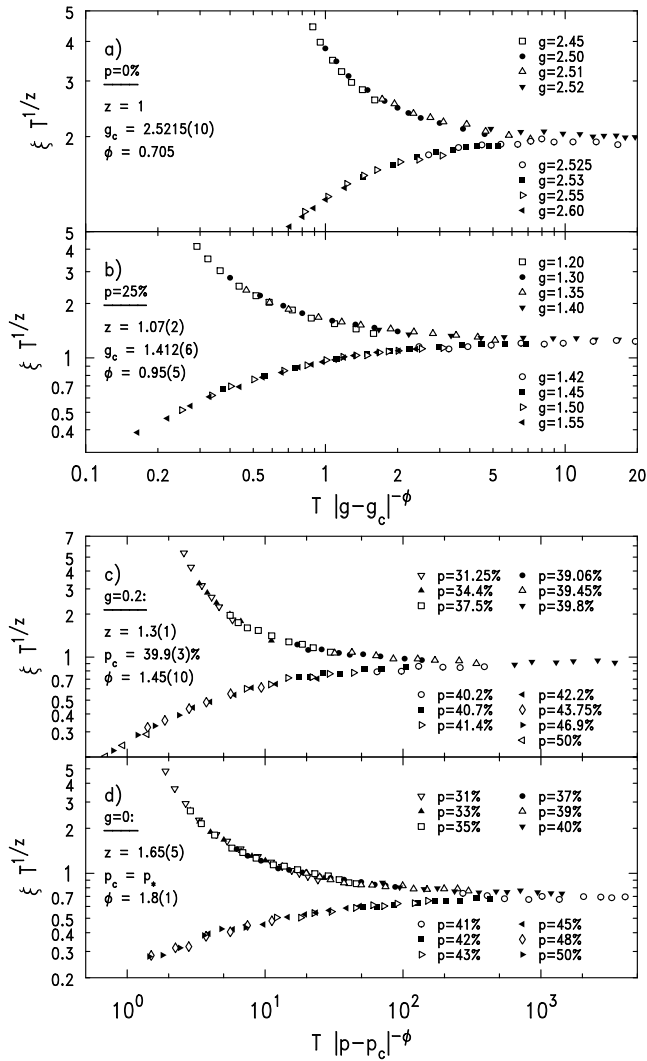}
\caption{\label{scaling}
    Scaling plots for the correlation length (a) for the pure spin-1/2 bilayer
    Heisenberg antiferromagnet, (b) at fixed nonmagnetic concentration
    $p=0.25$, and at fixed bilayer couplings (c) $g=0.20$ and (d) $g=0$ (SLHAF).
           }
\end{figure}

Sandvik \cite{sandvik02a}
concludes that the percolating cluster of the NN SLHAF at
$p=p_*$ has a nonzero ordered moment per spin, so that even in the extreme
quantum limit of spin-1/2 the percolation transition at $T=0$ is still
classical. If this picture is correct, then
one would expect the moment per spin on the critical cluster to
vanish at some nonzero value $g_m = g_c(p_*)$. This is indeed 
consistent with our findings. 
Figure 3 shows the
extrapolated ground state phase diagram, and we obtain
$g_m = 0.15(3)$.
The exponents for
$g=0.10$ are very similar to those at $g = 0.20$, and we estimate
$z_m = 1.33(7)$, $\nu_m = 1.11(14)$, and $\gamma_m = 2.69(31)$.
Calculations of the temperature dependence of the correlation length
at this multicritical point agree with this estimate for $z_m$.

The Harris criterion is often used to determine whether the
exponents at classical phase transitions should change in the
presence of weak disorder: for disorder to be relevant,
$\nu$ of the $pure$ system has to satisfy $\nu < 2/d$
\cite{harris74}.
More generally, it is expected that the $disordered$ system
satisfies $\nu \ge 2/d$ \cite{chayes86}. For a quantum phase
transition, the dimension to use is the
number of dimensions in which there is disorder, in our case
$d=2$. Since for the pure bilayer $\nu \approx 0.705 < 1$, we
expect disorder to be relevant \cite{sachdev92}.

\begin{figure}
\includegraphics[width=8.5cm]{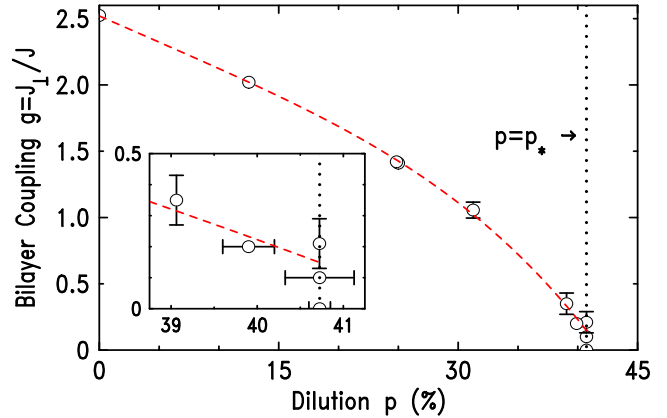}
\caption{\label{phase}
    Phase diagram of the diluted bilayer. The dashed
    line is a guide to the eye. The
    percolation threshold is marked by the dotted vertical line. The inset is
    a close-up view of the region near the percolation threshold.
        }
\end{figure}

The meaning of the exponents obtained along the phase boundary
($0<p<p_*$) is not entirely clear. As argued above,
disorder is expected to be a relevant perturbation of the pure system.
This should lead to a unique set of critical exponents for the
randomly-diluted bilayer Heisenberg antiferromagnet, different
from those at $g_c(0)$ and $g_m$, and from the classical
percolation exponents.  There are no special values of $p$ between
$p=0$ and $p=p_*$, and the exponents appear to vary continuously
along the phase boundary. This is qualitatively similar to the
two-dimensional Ising magnet, for which quenched disorder is a
relevant perturbation, and which appears to exhibit
(finite-temperature) transitions with concentration-dependent
exponents \cite{heuer92}. However, these Ising exponents also
show a slight temperature dependence, and it has
been argued that this indicates that these are not
the true exponents, which should only be observable at
temperatures asymptotically close to the transition temperature,
but rather effective exponents \cite{heuer92}.  A similar
masking might occur in the present system, so that true
critical exponents would only be revealed at asymptotically low
temperatures.  The critical points at $g_c(0)$ and $g_m$ influence
the intermediate-$T$ physics of systems nearby in parameter
space. One might expect that at $p \approx 25\%$ the
extracted exponents are close to the true critical exponents:
$p=25\%$ appears to be far away from both the pure system and from the
percolation threshold, so that disorder effects might
be large, while $\xi$ at the temperatures of our study
is still much larger than the percolation
length, which diverges at $p=p_*$. However, we obtain
$ \nu = \phi/z = 0.89$, in violation of the criterion
$\nu \ge 1$ \cite{chayes86}. Consequently, we believe that the exponents
extracted for $0 < p < p_*$ are effective, rather than true critical
exponents.

The scaling dimension for a quantum phase transition is $d_q = d+z$,
where $z$ can differ from 1 in the disordered case.
Using $z_m=1.33$, $\nu_m = 1.11$, and $\gamma_m = 2.69$,
the hyperscaling relationship $\beta = (d_q \nu - \gamma)/2$
gives $\beta_m \approx 0.5$
for the multicritical point.
Our corresponding effective values for $g=0$ are $z=1.65(5)$,
$\nu = 1.09(9)$, and $\gamma = 2.65(27)$,
and $\beta \approx 0.67(13)$.
Previous finite-size scaling results
for the NN SLHAF resulted in spin-dependent critical exponents, with
$z = 2.54(8)$, $\nu = 1.23(16)$ and $\beta = 0.50(7)$ for spin-1/2
\cite{kato00}.
Based on our results, these exponents have to be viewed as effective
exponents due to the influence of the nearby multicritical point.
This is consistent with the claim that the spin-1/2 NN SLHAF
at $p=p_*$ should exhibit asymptotic percolation critical behavior
\cite{sandvik02a}, for which
$\nu_p = 4/3 \approx 1.33$, $\gamma_p = 43/18 \approx 2.39$, and
$\beta_p = 5/36 \approx 0.14$ \cite{stauffer}.
Experiments for \lczmo~near $p=p_*$
revealed $\xi \sim T^{-\nu_T}$, with $\nu_T \approx
0.7$, which is very close to the effective value
$1/z \approx 0.61$ ($g=0$) and to
$1/z_m \approx 0.75$ ($g_m=0.15$).

In summary, we have mapped out the phase diagram of the
spin-1/2 bilayer Heisenberg antiferromagnet with quenched disorder
in form of interlayer dimer
dilution. Varying the bilayer coupling has allowed us to
further increase quantum fluctuations beyond those of the spin-1/2
NN SLHAF and to investigate the joint effects of quantum fluctuations
and quenched disorder. Our results for the pure bilayer agree with
previous work. We find that the critical coupling $g_c(p)$
decreases with increasing $p$, but remains nonzero even at the
percolation threshold $p=p_*$. The point $g_m = g_c(p_*)$ is a new
multicritical point, and we obtain estimates of several
critical exponents. Quenched disorder is expected to be
a relevant perturbation to the pure quantum system and to lead to
new critical behavior at nonzero $p$ below the percolation threshold.
The critical exponents along the phase boundary ($0< p < p_*$) appear
to change continuously, but it is likely that the true critical
behavior is masked by finite-temperature effects, and by the effective
proximity to either the pure fixed point or the new multicritical point.
The small value of $g_m \approx 0.15$ indicates that the
intermediate-temperature properties of the spin-1/2 SLHAF
\lczmo~near $p=p_*$ \cite{vajk02} may be
controlled by this nearby quantum critical point.
We note, that the experimental system might be best described by considering
a nonzero, frustrating
next-NN exchange of about $0.05-0.10 J$ \cite{kim01,vajk02},
which, in effect, would place it even closer to the multicritical point.
The phase diagram of Fig. 3 is qualitatively similar to that for the
randomly diluted 2D Ising model in a transverse field
\cite{senthil96}. We note that recent numerical work \cite{sandvik02b},
which uses a
different approach, gives $g_m(p_*) = 0.16(1)$ with $z_m = 1.28(2)$,
in very good agreement with our results.

We thank A. Aharony for suggesting the scaling analysis and A.W. Sandvik for
informing us of his numerical results.
This work was supported by the U.S. Department of
Energy under contract nos. DE-FG03-99ER45773 and DE-AC03-76SF00515, and by NSF
CAREER Award no. DMR9985067.

\bibliography{FINAL}       

\end{document}